\documentstyle[12pt]{article}
\begin{document}
\begin{center}

{\bf Scale Invariance and Vacuum Energy}
\end{center}
\bigskip
\begin{center}
\bigskip
E. I. Guendelman\footnote{Electronic address: guendel@bgumail.bgu.ac.il} 

{\it Physics Department, Ben Gurion University, Beer Sheva, Israel}
\end{center}

\bigskip

\begin{abstract}
The possibility of mass in the context of scale-invariant, generally
covariant theories, is discussed. Scale invariance is considered  in 
the context of a gravitational theory where
the action, in the first order formalism, is of the form $S =
\int L_{1} \Phi d^4x$ + $\int L_{2}\sqrt{-g}d^4x$ where $\Phi$ is a
density built out of degrees of freedom independent of the metric. 
For global scale invariance, a "dilaton"
$\phi$ has to be introduced, with non-trivial potentials $V(\phi)$ =
$f_{1}e^{\alpha\phi}$ in $L_1$ and $U(\phi)$ = $f_{2}e^{2\alpha\phi}$ in
$L_2$. This leads to non-trivial mass generation and a potential for
$\phi$ which is interesting for new inflation. Scale invariant mass terms 
for fermions lead to a possible explanation of the present day accelerated
universe and of cosmic coincidences.
\end{abstract}

The concept of scale invariance appears as an attractive possibility for a
fundamental symmetry of nature. In its most naive realizations, such a
symmetry is not a viable symmetry, however, since nature seems to have
chosen some typical scales.

Here we will find that scale invariance can nevertheless be incorporated
into realistic, generally covariant field theories. However, scale
invariance has to be discussed in a more general framework than that of
standard generally relativistic theories, where we must allow in the
action, in addition to the
ordinary measure of integration $\sqrt{-g}d^{4}x$, another one,
$\Phi d^{4}x$, where $\Phi$ is a density built out of degrees of freedom
independent of the metric.

	For example, given 4-scalars $\varphi_{a}$ (a = 
1,2,3,4), one can construct the density
\begin{equation} 
\Phi =  \varepsilon^{\mu\nu\alpha\beta}  \varepsilon_{abcd}
\partial_{\mu} \varphi_{a} \partial_{\nu} \varphi_{b} \partial_{\alpha}
\varphi_{c} \partial_{\beta} \varphi_{d}  
\end{equation}

	One can allow both geometrical
objects to enter the theory and consider$^1$
\begin{equation}
S = \int L_{1} \Phi  d^{4} x  +  \int L_{2} \sqrt{-g}d^{4}x		
\end{equation}

	 Here $L_{1}$ and $L_{2}$ are
$\varphi_{a}$  independent. There is a good reason not to consider 
mixing of  $\Phi$ and
$\sqrt{-g}$ , like
for example using
$\frac{\Phi^{2}}{\sqrt{-g}}$. This is because (2) is invariant (up to 
the integral of a total
divergence) under the infinite dimensional symmetry
$\varphi_{a} \rightarrow \varphi_{a}  +  f_{a} (L_{1})$
where $f_{a} (L_{1})$ is an arbitrary function of $L_{1}$ if $L_{1}$ and
$L_{2}$ are $\varphi_{a}$
independent. Such symmetry (up to the integral of a total divergence) is
absent if mixed terms are present.

	We will study now the dynamics of a scalar field $\phi$ interacting
with gravity as given by the action (2) with$^2$ 
\begin{equation}
L_{1} = \frac{-1}{\kappa} R(\Gamma, g) + \frac{1}{2} g^{\mu\nu}
\partial_{\mu} \phi \partial_{\nu} \phi - V(\phi),  L_{2} = U(\phi)
\end{equation}

\begin{equation}	
R(\Gamma,g) =  g^{\mu\nu}  R_{\mu\nu} (\Gamma) , R_{\mu\nu}
(\Gamma) = R^{\lambda}_{\mu\nu\lambda}, R^{\lambda}_{\mu\nu\sigma} (\Gamma) = \Gamma^{\lambda}_
{\mu\nu,\sigma} - \Gamma^{\lambda}_{\mu\sigma,\nu} +
\Gamma^{\lambda}_{\alpha\sigma}  \Gamma^{\alpha}_{\mu\nu} -
\Gamma^{\lambda}_{\alpha\nu} \Gamma^{\alpha}_{\mu\sigma}.	 
\end{equation}

	In the variational principle $\Gamma^{\lambda}_{\mu\nu},
g_{\mu\nu}$, the measure fields scalars
$\varphi_{a}$ and the  scalar field $\phi$ are all to be treated
as independent variables.

	If we perform the global scale transformation ($\theta$ =
constant) 
\begin{equation}
g_{\mu\nu}  \rightarrow   e^{\theta}  g_{\mu\nu}	
\end{equation}
then (2), with the definitions (3), (4), is invariant provided  $V(\phi)$ 
and $U(\phi)$ are of the
form  
\begin{equation}
V(\phi) = f_{1}  e^{\alpha\phi},  U(\phi) =  f_{2}
e^{2\alpha\phi}
\end{equation}
and $\varphi_{a}$ is transformed according to
$\varphi_{a}   \rightarrow   \lambda_{a} \varphi_{a}$
(no sum on a) which means
$\Phi \rightarrow \biggl(\prod_{a} {\lambda}_{a}\biggr) \Phi \equiv \lambda 
\Phi $
such that
$\lambda = e^{\theta}$
and	 
$\phi \rightarrow \phi - \frac{\theta}{\alpha}$. In this case we call the 
scalar field $\phi$ needed to implement scale invariance "dilaton".

	Let us consider the equations which are obtained from
the variation of the $\varphi_{a}$
fields. We obtain then  $A^{\mu}_{a} \partial_{\mu} L_{1} = 0$   	
where  $A^{\mu}_{a} = \varepsilon^{\mu\nu\alpha\beta}
\varepsilon_{abcd} \partial_{\nu} \varphi_{b} \partial_{\alpha}
\varphi_{c} \partial_{\beta} \varphi_{d}$. Since 
det $(A^{\mu}_{a}) =\frac{4^{-4}}{4!} \Phi^{3} \neq 0$ if $\Phi\neq 0$.
Therefore if $\Phi\neq 0$ we obtain that $\partial_{\mu} L_{1} = 0$,
 or that
$L_{1}  = M$, 	 
where M is constant. This constant M appears in a self-consistency 
condition of the equations of motion 
that allows us to solve for $ \chi \equiv \frac{\Phi}{\sqrt{-g}}$

\begin{equation}
\chi = \frac{2U(\phi)}{M+V(\phi)}.
\end{equation}

	To get the physical content of the theory, it is convenient to go
to the Einstein conformal frame where 
\begin{equation}
\overline{g}_{\mu\nu} = \chi g_{\mu\nu}		    
\end{equation}
and $\chi$  given by (7). In terms of $\overline{g}_{\mu\nu}$   the non
Riemannian contribution (defined   as                                
$\Sigma^{\lambda}_{\mu\nu} =                                                  
\Gamma^{\lambda}_{\mu\nu} -\{^{\lambda}_{\mu\nu}\}$                           
where $\{^{\lambda}_{\mu\nu}\}$   is the Christoffel symbol),
disappears from the equations, which can be written then in the Einstein
form ($R_{\mu\nu} (\overline{g}_{\alpha\beta})$ =  usual Ricci tensor)
\begin{equation}
R_{\mu\nu} (\overline{g}_{\alpha\beta}) - \frac{1}{2} 
\overline{g}_{\mu\nu}
R(\overline{g}_{\alpha\beta}) = \frac{\kappa}{2} T^{eff}_{\mu\nu}
(\phi)	 	
\end{equation}
where
\begin{equation}	 
T^{eff}_{\mu\nu} (\phi) = \phi_{,\mu} \phi_{,\nu} - \frac{1}{2} \overline
{g}_{\mu\nu} \phi_{,\alpha} \phi_{,\beta} \overline{g}^{\alpha\beta}
+ \overline{g}_{\mu\nu} V_{eff} (\phi),
V_{eff} (\phi) = \frac{1}{4U(\phi)}  (V+M)^{2}.
\end{equation}
	If $V(\phi) = f_{1} e^{\alpha\phi}$  and  $U(\phi) = f_{2}
e^{2\alpha\phi}$ as
required by scale invariance, we obtain from (10)
\begin{equation}
	V_{eff}  = \frac{1}{4f_{2}}  (f_{1}  +  M e^{-\alpha\phi})^{2}	
\end{equation}

	Since we can always perform the transformation $\phi \rightarrow
- \phi$ we can
choose by convention $\alpha > O$. We then see that as $\phi \rightarrow
\infty, V_{eff} \rightarrow \frac{f_{1}^{2}}{4f_{2}} =$ const.
providing an infinite flat region. Also a minimum is achieved at zero
cosmological constant for the case $\frac{f_{1}}{M} < O$ at the point 
$\phi_{min}  =  \frac{-1}{\alpha} ln \mid\frac{f_1}{M}\mid $. Finally, 
the second derivative of the potential  $V_{eff}$  at the minimum is 
$V^{\prime\prime}_{eff} = \frac{\alpha^2}{2f_2} \mid{f_1}\mid^{2} > O$
if $f_{2} > O$,	 	

There are many interesting issues that one can raise here. The first one
is of course the fact that a realistic scalar field potential, with
massive excitations when considering the true vacuum state, is achieved in
a way consistent with the idea of scale invariance.

	The second point to be raised is that since there is an infinite 
region of flat potential for $\phi \rightarrow \infty$, we expect a slow 
rolling new
inflationary$^3$ scenario to be
viable, provided the universe is started at a sufficiently large value of
the scalar field $\phi$.

            Furthermore, one can consider this model as suitable for the
present day universe rather than for the early universe, after we suitably
reinterpret the meaning of the scalar field  $\phi$. This can provide a long 
lived almost constant vacuum energy for a
long period of time, which can be small if $f_{1}^{2}/4f_{2}$ is
small. Such small energy
density will eventually disappear when the universe achieves its true
vacuum state. 

	Notice that a small value of $\frac{f_{1}^{2}}{f_{2}}$   can be
achieved if we let $f_{2} >> f_{1}$. In this case
$\frac{f_{1}^{2}}{f_{2}} << f_{1}$, i.e. a very small scale for the
energy
density of the universe is obtained by the existence of a very high scale
(that of $f_{2}$) the same way as a small fermion mass is obtained in the
see-saw mechanism$^{4}$ from the existence also of a large mass scale. In 
what follows, we will take $f_{2} >> f_{1}$.

        So far we have studied a theory which contains the metric tensor
$g_{\mu\nu}$, the measure fields $\varphi_{a}$ (a=1,2,3,4) and the
"dilaton" $\phi$, which makes global scale invariance possible in a
non-trivial way. All of the above fields have some kind of geometrical
significance, but if we are to describe the real world, the list of fields
and/or particles has to be enlarged.

Taking, for example, the case of a fermion $\psi$, where the kinetic term
of the fermion is chosen to be part of $L_1$
\begin{equation}
S_{fk} = \int L_{fk} \Phi d^4 x
\end{equation}
\begin{equation}
L_{fk} = \frac{i}{2} \overline{\psi} [\gamma^a V_a^\mu
(\overrightarrow{\partial}_\mu + \frac{1}{2} \omega_\mu^{cd} \sigma_{cd})
- (\overleftarrow{\partial}_\mu + \frac{1}{2} \omega_\mu^{cd} \sigma_{cd})
\gamma^a V^\mu_a] \psi
\end{equation}
there $V^\mu_a$ is the vierbein, $\sigma_{cd}$ =
$\frac{1}{2}[\gamma_c,\gamma_d]$, the spin connection $\omega^{cd}_\mu$ is
determined by variation with respect to $\omega^{cd}_\mu$ and, for
self-consistency, the curvature scalar is taken to be (if we want to deal
with $\omega_\mu^{ab}$ instead of $\Gamma^\lambda_{\mu\nu}$ everywhere)
\begin{equation}
R = V^{a\mu}V^{b\nu}R_{\mu\nu ab}(\omega),
R_{\mu\nu ab}(\omega)=\partial_{\mu}\omega_{\nu ab}
-\partial_{\nu}\omega_{\mu ab}+(\omega_{\mu a}^{c}\omega_{\nu cb}
-\omega_{\nu a}^{c}\omega_{\mu cb}).
\end{equation}

Global scale invariance is obtained
provided $\psi$ also transforms, as in
$\psi \rightarrow \lambda ^{-\frac{1}{4}} \psi$. Mass term consistent with 
scale invariance exist,
\begin{equation}
S_{fm} = m_1 \int \overline{\psi} \psi e^{\alpha\phi/2} \Phi d^4x + m_2
\int \overline{\psi} \psi e^{3\alpha\phi/2} \sqrt{-g} d^4 x. 
\end{equation}

If we consider the situation where 
$m_1 e^{\alpha\phi/2} \overline{\psi}\psi$  
or $m_2 e^{3\alpha\phi/2} \overline{\psi}\psi$ are
much bigger than $V(\phi)$ + M, i.e. a high density approximation, we
obtain that instead of (7) that the consistency condition is$^1$ 
 $(3m_2 e^{3\alpha\phi/2} + m_1 e ^{\alpha\phi/2} \chi) \overline{\psi}
\psi = 0$,
which means
$\chi = -\frac{3m_2}{m_1} e^{\alpha\phi}$. Using this in (15), we obtain,  
after going to the conformal Einstein frame, which involves, 
also a transformation of the fermion fields,
necessary so as to achieve  
Einstein-Cartan form for both the gravitational and fermion 
equations. These transformations are, 
$\overline{g}_{\mu\nu}$ = $\chi g_{\mu\nu}$ (or
$\overline V_\mu^a$ = $\chi^\frac{1}{2} V_\mu^a$) and $\psi ^\prime$ =
$\chi ^{-\frac{1}{4}} \psi$ and they lead to a mass term,
\begin{equation}
S_{fm} = -2m_2 ( \frac {|m_1|}{3|m_2|})^{3/2}  \int\sqrt{-\overline {g}} 
\overline{\psi} ^{\prime} \psi ^{\prime} d^4x
\end{equation}

The $\phi$ dependence of the mass term has disappeared, i.e. masses are 
constants.

There is one situation where the low density of matter can also give results 
which are similar to those obtained in the high density approximation, in 
that the coupling of the $ \phi $ field disappears and that the mass term 
becomes of a conventional form in the Einstein conformal frame. 

This is the case, when we study the theory for the limit 
$\phi \rightarrow \infty$ . Then $U(\phi) \rightarrow \infty$ and
$V(\phi) \rightarrow \infty$. In this case, taking
$m_1 e^{\alpha\phi/2} \overline{\psi}\psi$                                    
and $m_2 e^{3\alpha\phi/2} \overline{\psi}\psi$                                                                            
much smaller than $V(\phi)$ or $U(\phi)$ respectively, therefore one can see 
that (7) is a good approximation and since 
also $M$ can be ignored in the self consistency condition (7)
in this limit, we get then, 
$\chi = \frac{2f_2}{f_1} e^{\alpha\phi}$. If this is inserted in (15), 
we get $S_{fm} = m \int\sqrt{-\overline {g}} 
\overline{\psi} ^{\prime} \psi ^{\prime} d^4x$, where

\begin{equation}
 m = m_1(\frac {f_1}{2f_2})^{\frac{1}{2}} +  m_2(\frac {f_1}{2f_2})^{\frac{3}{2}}
\end{equation}

Comparing (16) and (17) and taking  $m_1$ and $m_2$ of the same 
order of magnitude, we see that the mass of the Dirac particle is much 
smaller in the region $\phi \rightarrow \infty$, for which (17) is 
valid, than it is in the region of high density of the Dirac particle 
relative to $V(\phi)+M$, as displayed in eq. (16), if the "see-saw" 
assumption $\frac{f_1}{f_2} < < 1$ is made.

Therefore if space is populated by these diluted Dirac particles of this
type, the mass of these particles will grow substantially if we go to the true vacuum state 
valid in the absence of matter, i.e. $V+M=0$, as dictated by $V_{eff}$ given 
by eq. (11).
The presence of matter pushes therefore the minimum of energy to a state 
where $ V+M > 0$. The real vacuum in the presence of matter should not 
be located in the region $\phi \rightarrow \infty$, which minimizes the 
matter energy, but maximizes the potential energy $V_{eff}$ and not at
$V+M=0$, which minimizes $V_{eff}$, and where particle masses are big, but
somewhere in a balanced intermediate stage. Clearly how much above $V+M=0$
such true vacuum is located must be correlated to how much particle density 
is there in the Universe. A non zero vacuum energy, which must be of the
same order of the particle energy density, has to appear and this could 
explain the "accelerated universe" that appears to be implied by the most 
recent observations, together with the "cosmic coincidence", that requires 
the vacuum energy be of the same order of magnitude to the matter energy$^5$.

I would like to thank J. Bekenstein, A. Davidson, A.Guth and 
A. Kaganovich for conversations on the subjects discussed here.

\end{document}